\documentclass[twocolumn]{article}

\usepackage{array}
\usepackage{authblk}
\usepackage{hyperref}
\usepackage[margin=2cm]{geometry}
\usepackage{graphicx}
\usepackage{dcolumn}
\usepackage{bm}
\usepackage{authblk}
\usepackage{amsmath}

\newcommand{\rr}{{\bf r}}

\usepackage{mathptmx}

\newcommand{\rep}{{\rm{rep}}}
\newcommand{\xx}{\mathbf{x}}

\providecommand{\keywords}[1]
{
  \small	
  \textbf{\textit{Keywords---}} #1
}

\begin{document}

\title{Generalised Kohn-Sham equations with accurate total energy and single-particle eigenvalue spectrum}
\author{Thomas C. Pitts$^1$, Nektarios N. Lathiotakis$^2$, Nikitas I. Gidopoulos$^1$ \\
\small $^1$Department of Physics, Durham University, South Road, Durham, DH1 3LE, UK \\
$^2$ Theoretical and Physical Chemistry Institute, National Hellenic Research Foundation, Vass. Constantinou 48, Athens GR-11635, Greece.
}
\date{\today}

\twocolumn[
  \begin{@twocolumnfalse}
\maketitle

\begin{abstract}
We propose a new generalised Kohn-Sham or constrained hybrid method, where the exchange potential is the (equally weighted) 
average of the nonlocal Fock exchange term and the self-interaction-corrected exchange potential, as obtained from our constrained minimisation method of semi-local approximations.   
The new method gives an accurate single-particle eigenvalue spectrum with an average deviation between (the negative of) 
the valence orbital eigenvalues and the experimental ionisation potentials of about 0.5eV, while the deviation of core orbitals is within 2eV. 
The improvement in the eigenvalue spectrum is achieved with a minimal increase in the total energy. 	 
\end{abstract}

 \\

\end{@twocolumnfalse}
]


\section{Introduction}

Historically, the Kohn-Sham (KS) scheme was proposed to provide an improved approximation for the kinetic energy of an electronic system in its 
ground state, as a functional of the density. 
As such, the (exact) KS equations were not expected to yield an accurate single-particle eigenvalue spectrum, except for the highest occupied 
(HOMO) eigenvalue which is equal to the ionisation potential (IP) of the system \cite{perdew1982density}. 

This view has changed over the years. Numerical inversion of accurate electronic densities to obtain a nearly exact KS potential 
has revealed that the eigenvalues of the valence KS orbitals are very close in magnitude (within $\sim$0.1eV) to the corresponding vertical IPs of the system. 
Bartlett and coworkers \cite{verma2012increasing,bartlett2019adventures} and Baerends and coworkers \cite{van2014physical} 
have independently proven theorems predicting the observed closeness 
of the negative of the eigenvalues of valence KS orbitals to the IPs of the electronic system. For deeper valence electrons the coincidence
between the negative of KS orbital eigenvalues and IPs deteriorates and for core electrons the difference reaches a few tens of electron volts.

Approximate local and semi-local density functionals yield rather accurate total energy predictions but their corresponding KS potentials are 
not equally good. 
The typical KS HOMO eigenvalue error of local and semi-local density functional approximations (DFAs) is about $\sim$4 eV, relative to the 
exact KS result. 
The reason is simple: for an $N$-electron system, the electrons in the occupied KS orbitals 
are repelled via the Hartree-exchange-correlation (HXC) part of the approximate KS potential 
by an effective charge of $N$ electrons, rather than the correct $N-1$ electrons of a self-interaction-free model. 
Consequently, the occupied orbital eigenvalues are erroneously shifted to higher energies, 
resulting in an underestimation of the IP by approximately $\sim$4 eVs.

In our group, we have addressed this error with a constrained minimisation of the total energy in local or semi-local DFAs.
The DFA total energy remains unchanged but the constrained minimisation forces the 
effective ``screening'' (or ``electron repulsion'') charge of the HXC potential of the DFA to equal $N-1$ 
\cite{gidopoulos2012constraining}. 
As a result, the negative of the HOMO eigenvalues from our constrained minimisation method come closer to the true IPs 
(still underestimating them on average) with a typical error of about $\sim$1 eV. It is noteworthy that the resulting IP predictions 
are similar for the three different DFAs we tried (LDA, PBE, B3LYP). 
\footnote{Using a local multiplicative KS potential for all three, including B3LYP.}

In Hartree Fock (HF) theory, Koopmans theorem predicts that the negative of the eigenvalues of all the occupied orbitals are equal to minus the 
corresponding vertical IPs. The theorem neglects orbital relaxation when an electron is ionised and HF theory omits electron correlation altogether. 
These two errors have opposite signs for occupied orbitals \cite{evertjahn2013} and almost cancel each other. 
Still, the orbital relaxation error is stronger than the correlation error and the dominance of relaxation increases for core electrons. 
The HF eigenvalues typically overestimate the vertical IPs from about $\sim$1 eV for the least bound electron (HOMO) 
to a few tens of eVs for core electrons.

\begin{figure*}
\begin{centering}
\includegraphics[width=0.9\linewidth]{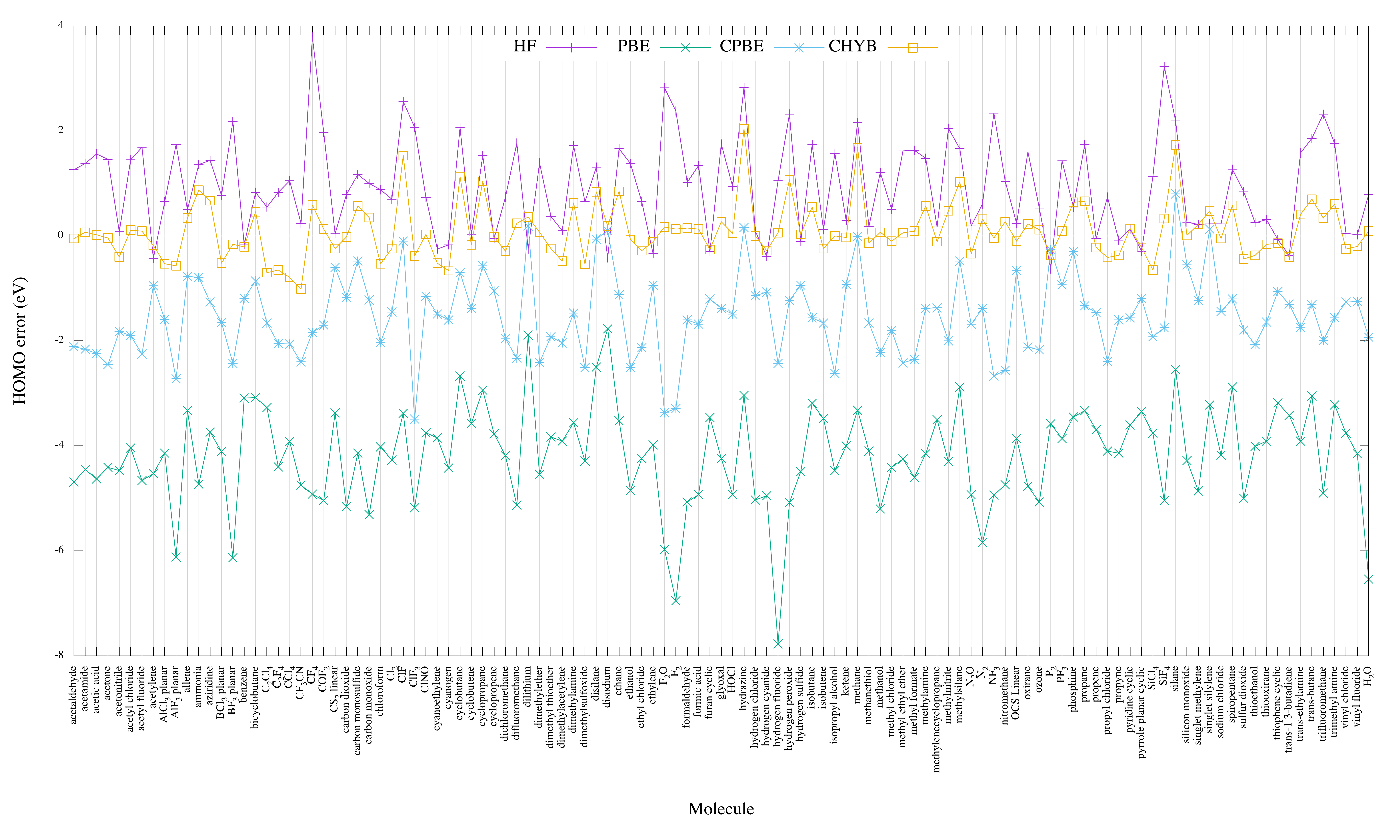} 
\caption{IP errors compared to experimental results for a number of molecules using a cc-pVDZ basis set. 
The theoretical IP values are obtained from the negative of the HOMO eigenvalues in CLDA/HF equations.} 
\end{centering}
\label{HF_LDA_comp}
\end{figure*}

The improvement of the HOMO eigenvalue with the constrained minimisation of local/semi-local DFAs brings the magnitude of the 
IP error to the same ballpark as HF theory, but with the opposite sign (Fig.~\ref{HF_LDA_comp}). 
At the same time, the total energies from DFA constrained minimisations 
are virtually identical to the unconstrained DFA energies (within $\sim$0.1 meV)\cite{pitts2018performance}, 
so the good quality of the DFA total energies remains intact in the 
constrained minimisation.    
The aim of this paper is to consider a hybrid of the constrained minimisation of local/semi-local DFAs and of HF theory, aiming to 
formulate a generalised KS (GKS) scheme with further improved single-particle spectrum over constrained DFA (CDFA) 
and HF theory for all occupied electrons, preserving at the same time the quality of the DFA total energy.

The rest of the paper is organised as follows: in section II we review briefly the CDFA method and modify it to aid its integration 
in a hybrid scheme and to make the implementation of our constraints more efficient. 
In section III we present the GKS formalism valid for any mixing between CDFA and HF. 
We seek the optimal mixing and find it is close to 50\%, as expected from Fig.~\ref{HF_LDA_comp}. 
In section IV we confirm that the predictions for the first IPs are better than those of either HF or CDFA, 
and that the quality of the resulting total energies is of similar quality compared with those of the DFA (and CDFA). 
Finally, we investigate whether the flexibility of our hybrid potential, corrected for self-interactions (SIs) 
allows for a better prediction for the IPs of all the bound electrons, including those in the core 
where relaxation effects are expected to become dominant.   

\section{Constrained minimisation of the total energy}

In our previous work\cite{gidopoulos2012constraining,pitts2018performance} the approximate
KS potential is constrained in order to remove the effect of SIs from its long-range asymptotic decay 
(without correcting for SIs the total energy).
For a finite system of $N$ electrons, the HXC potential should decay as $(N-1)/r$, but 
in many common approximations it incorrectly decays  as $N/r$, a failure that we attribute to SIs. 

To correct the potential for SIs the KS equations are written as 
\begin{equation}
\left[ -\frac{\nabla^2}{2} + v_{en}(\rr) + v_{\rep}(\rr) \right] \phi_i(\rr ) = \epsilon_i\phi_i(\rr ) ,
\end{equation}
where $v_{\rep}$ takes the role of the HXC potential, $v_{\rm{HXC}}$. 
The potential $v_{\rep}$ is written as the Coulomb Hartree potential of an auxiliary density $\rho_{\rep}$:
\begin{equation} 
v_{\rep}(\rr )  = \int d\rr'  \frac{\rho_{\rep}(\rr' ) }{|\rr'  - \rr |} .
\end{equation}
We have called this density the ``effective repulsion'' or ``screening'' density in the KS system. 
Mimicking in a mean field way the absent repulsion of the real electrons, the KS screening density is 
distinct from the physical density of the interacting electron system and is subject to the following constraints:
\begin{eqnarray}
\int d\rr  \rho_{\rep}(\rr ) &= N-1 , \label{constraint1}\\
\rho_{\rep}(\rr ) &\geq 0 . \label{constraint2}
\end{eqnarray}
The application of these constraints ensures that the potential $v_{\rep}$ has the correct asymptotic decay of the exact HXC potential. 
Additionally, for a single electron the effective repulsion potential becomes zero everywhere,  $v_{\rep} (\rr) =0$, which is 
the exact one electron result for the HXC potential. 
Even though we employ $v_{\rm rep}$ to approximate $v_{\rm HXC}$, we avoid writing $v_{\rm rep} = v_{\rm HXC}$ because so far, 
we have not been able to write $v_{\rm rep}$ satisfying (\ref{constraint1},~\ref{constraint2}) 
as the functional derivative of an approximate HXC energy.

The first constraint in (\ref{constraint1}) is incorporated via a Lagrange multiplier in the objective functional
\begin{equation}\label{objfunc}
G^{\rm{DFA}}[\rho_\rep] = E^{\rm{DFA}}[\rho_\rep] -\lambda \int d\rr  \rho_{\rep}(\rr ) .
\end{equation}
In Refs. \cite{gidopoulos2012constraining,pitts2018performance} the second constraint in (\ref{constraint2}) is enforced through the use of 
a penalty function. 
However, our recent work\cite{EFOfuturework} provides an alternative, 
by constructing the density $\rho_{\rep}(\rr )$ as the modulus 
square of an effective repulsion, or screening, amplitude, 
\begin{equation} \label{EFOmethod}
\rho_{\rep}(\rr )= |f(\rr )|^2 .
\end{equation}

To obtain the ground state, the objective functional in Eq. \ref{objfunc} is minimised with respect to $\rho_{\rep}$. 
The  minimisation can be performed as an extension to the optimised effective potential (OEP) 
method\cite{sharp1953variational,talman1976optimized}.
(When $\rho_{\rep}$ is given by Eq.~\ref{EFOmethod}, the minimisation is with respect to $f$. 
Details of the minimisation of the objective function with respect to the amplitude $f$ in 
(\ref{objfunc}) are presented in a separate publication.) 
At the minimum of $G$,
\begin{multline}
\label{original_min}
0 = \frac{\delta G^{\rm{DFA}}[\rho_{\rep}]}{\delta \rho_{\rep}(\xx) }  = \\ 
\int d\rr d\rr'  \frac{\delta v_{\rep}(\rr')}{\delta \rho_\rep(\xx)}  \chi_v(\rr , \rr' ) 
[ v^{\rm DFA}_{\rm{HXC}}(\rr) - v_{\rep}(\rr) ] -  \lambda  , 
\end{multline} 
where the potential $v^{\rm DFA}_{\rm{HXC}}(\rr)={\delta E^{\rm DFA}_{\rm{HXC}}[\rho]}/{\delta \rho(\rr)}$ is the DFA HXC 
potential and
\begin{equation}
\chi_v(\rr , \rr' ) = -2 \sum_{ia} \frac{\phi_i(\rr ) \phi_a(\rr ) \phi_i(\rr' ) \phi_a(\rr' )}{\epsilon_a - \epsilon_i}
\end{equation}
is the KS density-density response function. 

This method can be applied to a generic energy density functional $E[\rho]$. 
In Ref.~\cite{pitts2018performance} we applied this method to a range of functionals, including a hybrid functional for which 
the HXC potential is non-local. The constrained minimisation increased total energies minimally, 
while improving calculations of the IP from the (negative) of the HOMO eigenvalue. 
It was also found that the HOMO eigenvalues were largely independent from the functional used to calculate them. 

In the original formulation of the constrained method the sum of the Hartree, 
exchange and correlation potentials is subject to the constraints (\ref{constraint1}\ref{constraint2}). 
Alternatively, these constraints can instead be applied only to the Hartree and exchange (HX) potential. 
To leading order the long range decay of the exact HX potential is $(N-1)/r$ 
while the exact correlation potential decays as $\propto -1/r^4$\cite{van1994exchange,almbladh1984density}. 
Therefore the asymptotic behaviour of the exact $v_{\rm{HXC}}$ will be dominated by that of the HX potential $v_{\rm{HX}}$. 
The exact, leading-order asymptotic behaviour of the potential is therefore maintained if the unconstrained correlation potential is included 
separately to the constrained HX potential.
Correlation can be treated separately to the constrained method, by constructing the KS equations as
\begin{equation} 
\left[ -\frac{\nabla^2}{2} + v_{en}(\rr ) + v_{\rm c}(\rr ) + v_{\rep}(\rr ) \right] \phi_i(\rr ) = \epsilon_i\phi_i(\rr )
\end{equation}
where $v_{\rm c} (\rr )$ is the DFT correlation potential ${\delta E_{\rm c}[\rho]}/{\delta \rho(\rr )}$.
Treating correlation separately to Hartree-exchange, represented again by $v_{\rm rep}$, 
the equation to determine $\rho_\rep$ becomes, 
\begin{multline}
\label{new_min}
0 = \frac{\delta G^{\rm{DFA}}[\rho_{\rep}]}{\delta \rho_{\rep}(\xx) }  = \\
\int d\rr d\rr'  \frac{\delta v_{\rep}(\rr')}{\delta \rho_\rep(\xx)}  \chi_v(\rr , \rr' ) 
[ v^{\rm{DFA}}_{\rm{HX}}(\rr) - v_{\rep}(\rr) ] 
-  \lambda ,
\end{multline}
identical to Eq.~\ref{original_min} except the potential $v_{\rm{HXC}}^{\rm DFA}$ is now $v_{\rm{HX}}^{\rm DFA}$.
\setlength{\tabcolsep}{5pt}
\begin{table}
\begin{center}
\begin{tabular}{ccccccc}
\hline
\hline
\multicolumn{2}{c}{} & \multicolumn{2}{c}{LDA} & &\multicolumn{2}{c}{PBE} \\
\cline{3-4}\cline{6-7}
Molecule & Expt& $v_{\rm xc}$&  $v_{\rm x}$&&  $v_{\rm xc}$&  $v_{\rm x}$\\
\hline
C\textsubscript{2}H\textsubscript{2}&11.40&-1.52&-0.74&&-1.30&-1.03\\
NH\textsubscript{3}&10.07&-0.92&-0.14&&-0.84&-0.65\\
CO&14.01&-2.23&-1.22&&-1.71&-1.24\\
C\textsubscript{2}H\textsubscript{4}&10.51&-1.34&-0.57&&-1.28&-0.96\\
F\textsubscript{2}&15.70&-4.79&-3.50&&-4.07&-3.29\\
CH\textsubscript{2}O&10.89&-2.31&-1.44&&-2.04&-1.64\\
HCN&13.60&-1.85&-1.03&&-1.84&-1.32\\
HF&16.03&-2.82&-1.89&&-2.57&-2.06\\
CH\textsubscript{4}&12.61&-0.10&0.66&&0.01&0.31\\
N\textsubscript{2}&15.58&-2.82&-1.74&&-2.25&-1.71\\
H\textsubscript{2}O&12.62&-2.10&-1.30&&-1.97&-1.69\\
\hline 
&AE&-2.07&-1.17&&-1.81&-1.39 \\
&AAE&2.07&1.29&&1.81&1.45\\
&$\Delta E$&&$7.9 \times$&&&$-3.3 \times$ \\
&&&$10^{-6}$&&&$ 10^{-5}$ \\
\hline
\hline
\end{tabular} 
\caption{ The errors in calculating HOMO energies for a sample of molecules using the constrained minimisation method, with correlation included
 $v_{\rm xc}$ and correlation separate $v_{\rm x}$. 
Calculations performed for a cc-pVTZ basis set. The total average error (AE) and absolute average error (AAE) are shown along with the average change in energy between the two methods $\Delta E$. All energies in eVs. }
\end{center}
\label{Correl_results} 
\end{table}
Table \ref{Correl_results} shows that excluding correlation from the constraint results in an upshift of the calculated IP.
For LDA this results in an average upshift of $0.90$ eV, and for PBE this average is $0.42$ eV. 
%
%
Typically calculations of the HOMO energy, similarly to unconstrained LDA calculations, underestimate the IP. 
This upshift of the calculated IP therefore, after leaving correlation unconstrained, 
generally improves slightly the quality of the approximation for the IP of the system. 
However, the removal of correlation from the constrained minimisation results 
in a KS potential that is no longer exact for a system with a single electron. 
The exact behaviour for a one electron system is only recovered 
if the approximate correlation potential is zero for a single electron, which holds for the exact functional. 

\section{The Generalised KS Method} 

Separating the correlation potential from the constrained minimisation 
allows the calculation of a constrained HX component of the KS potential where the main effects of SIs have been removed. 
To construct our hybrid method, this potential is combined with the SI-free non-local HF potential in a GKS scheme. 
A simple generalised KS equation can be constructed using a single parameter 
$\alpha$ to control the contribution of local and non-local exchange
\begin{equation}\label{HYB_formula}
v_{\rm hyb} = (1-\alpha)  v_{\rm x}^{\rm F} + \alpha \, v^{\rm{DFA}}_{\rm eff} + v^{\rm{DFA}}_{\rm c}
\end{equation}
where $v_{\rm x}^{\rm F} $ is the nonlocal Fock exchange potential, 
$v_{\rm c}^{\rm DFA}$ is the DFA correlation potential and $v_{\rm eff}$ is the self-interaction corrected exchange potential,
(difference of $v_{\rm rep}$ and Hartree potentials, $v_{\rm eff} = v_{\rm rep} - v_{\rm H}$), with 
$v_{\rm rep}$ given by Eq. \ref{new_min}. 
Correspondingly the total energy is given as 
\begin{multline}
\label{total_energy}
E_{\rm tot}[\rho] = T_{\rm s}[\rho] + E_{\rm en}[\rho] + E_{\rm H}[\rho] + E^{\rm{DFA}}_{\rm c} \\
+ \alpha E^{\rm{DFA}}_{\rm x}[\rho] + (1-\alpha) E^{\rm F}_{\rm x} [\rho] 
\end{multline}
where $T_{\rm s}$ is the non-interacting kinetic energy, $E_{\rm en}$ is the energy from the nuclear potential, $E_{\rm H}$ is the Hartree energy,
 $E_{\rm x}^{\rm DFA}$ and $E_{\rm c}^{\rm DFA}$ are the exchange and correlation energy functionals, 
 and $E^{\rm F}_{\rm x}$ is the Fock-exchange energy.

The treatment of correlation separately to exchange allows for the new 
hybrid method to have a consistent correlation energy for all values of $\alpha$. 
\begin{table}
\begin{center}
\begin{tabular}{ccc}
\hline
\hline
$\alpha$&LDA&PBE\\
\hline
0.2&0.224&0.005\\
0.4&0.446&0.008\\
0.6&0.666&0.010\\
0.8&0.884&0.010\\
1&1.100&0.009\\\hline
\hline
\end{tabular} 
\end{center}
\caption{Table showing the change to the total energies, $E(\alpha) - E(\alpha=0)$, 
for various values of $\alpha$, for the LDA and PBE functionals and a range of molecules. 
These results are the average of the differences in total energies for the molecules in Table \ref{Correl_results}. Energies in eV.}
\label{corel_energies} 
\end{table}
The total energy varies almost linearly from the HF plus LDA correlation limit ($\alpha=0$), 
to the constrained DFA energy limit ($\alpha = 1$), as the hybrid parameter $\alpha$ varies from 0 to 1. 
We look to choose a DFA for which this change in total energy is small, such that total energies are insensitive to the choice of $\alpha$. 
This allows the parameter $\alpha$ to be varied to optimise the prediction of ionisation energies for this hybrid. 
For LDA, Table \ref{corel_energies} demonstrates a significant change of $1.11$ eV in the total energy as $\alpha$ varies over $0\leq\alpha\leq1$. 
This large energy change can be explained because the HF energy is lower than the LDA total energy and  
adding correlation energy, a negative quantity, on top of HF at $\alpha = 0$ lowers further the total energy and increases 
the energy difference between the end points, $\alpha = 1$ (LDA) and $\alpha = 0$ (HF+LDA correlation).  

The constrained PBE (CPBE) approximation shows a smaller energy shift of about $0.01$ eV (Table~\ref{corel_energies}) 
between the constrained PBE total energy ($\alpha=1$) and the HF plus PBE correlation energy ($\alpha=0$). 
To exploit the insensitivity of the total energy on the hybrid parameter $\alpha$ we restricted our investigation 
to the hybrid method with constrained PBE exchange. 
Because the energetics of the CPBE hybrid remain consistent throughout the range $0\leq\alpha\leq 1$, the effect of varying 
$\alpha$ on the orbital energies can be investigated while the quality of the energetics is not affected. 
As we found in our previous work\cite{pitts2018performance} the ionisation energy calculated from the constrained method is insensitive to the choice of DFA and therefore an optimisation of the ionisation energies is likely to be valid for a range of exchange functionals.  
%
\section{Results} 

\begin{figure}
\begin{centering}
\includegraphics[width=0.9\linewidth]{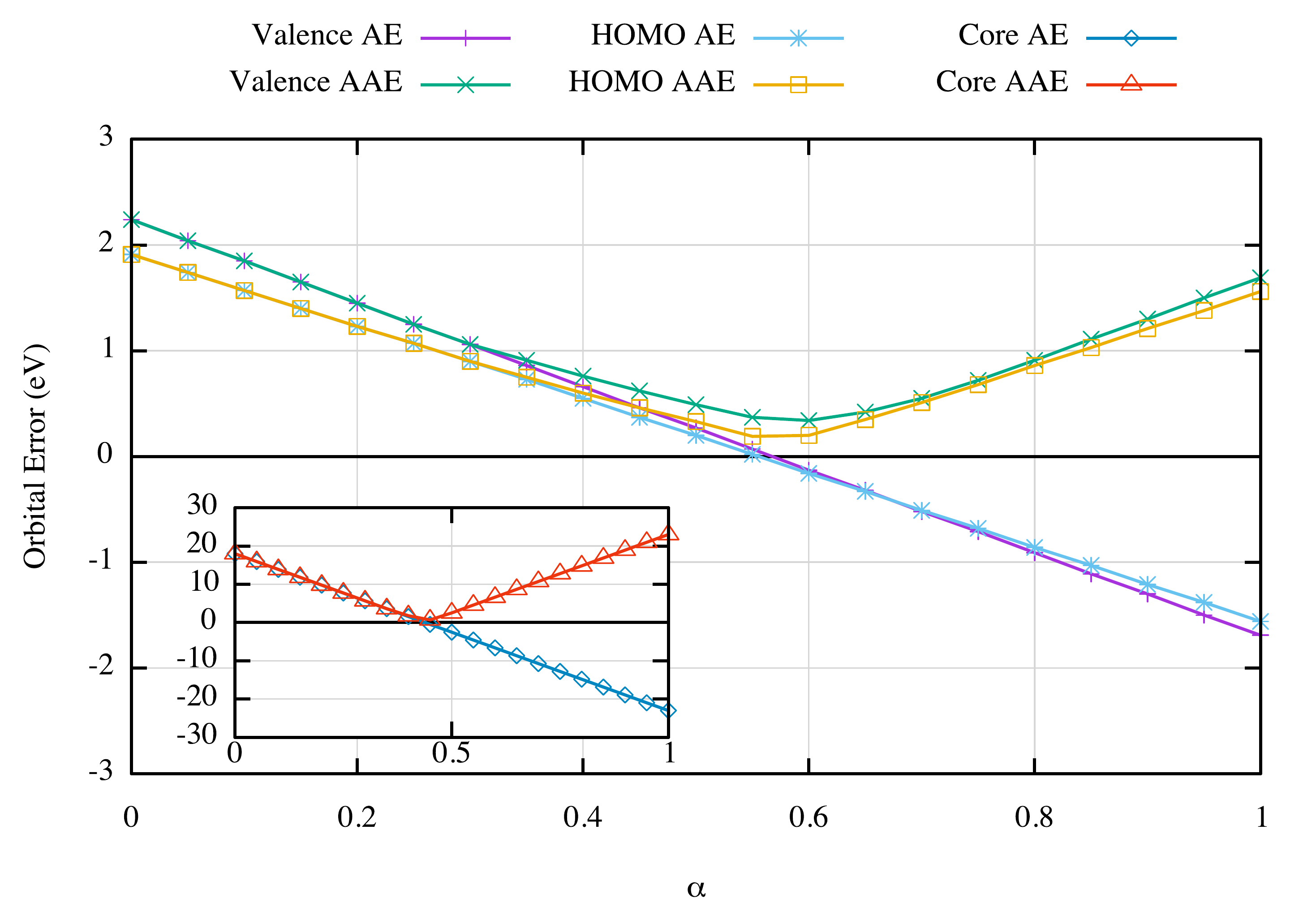} 
\caption{The orbital energy ``error'' (difference of orbital eigenvalue magnitude from experimental IP) as 
a function of $\alpha$. 
We show average error (AE) and absolute average error (AAE) for all the valence orbitals and separately for 
the HOMO. In the inset, we show 
the average error and absolute average error for the core orbitals. 
Results are calculated for the molecules in Table~\ref{Correl_results} using a cc-pVDZ basis set.  }
\end{centering}
\label{alpha_investigation}
\end{figure}

With our choice of CPBE, the total energies of our hybrid scheme are relatively unchanged for a range of values of $\alpha$. 
This allows the optimisation of the hybrid parameter $\alpha$ for the calculation of ionisation energies.
However, instead of focusing just on the HOMO eigenvalue\cite{van2014physical,doi:10.1063/1.4755818}, 
we optimised the present hybrid method to best approximate the ionisation energies of all occupied orbitals.  
The vaue of the hybrid parameter was optimised for the simple molecules in 
Table~\ref{Correl_results} with molecular geometries taken from the supplementary material of 
Ref.~\cite{ranasinghe2019vertical} along with experimental and coupled cluster results. 
The minimising $\alpha$ can be determined from the results in Fig. \ref{alpha_investigation}. 
These results show that the minimising $\alpha$ varies slightly when considering the HOMO, core, and valence orbitals. 
The energy eigenvalues of the valence orbitals are optimised at a slightly larger value of $\alpha$, i.e. a smaller percentage of HF exchange, 
while IPs from core orbital eigenvalues are better approximated with a slightly smaller percentage of constrained DFT exchange. 
A choice that achieves good accuracy for the full range of orbitals without overfitting the results is the value $\alpha=0.5$. 
This value coincides with the half-and-half mixing of the original hybrid formulation of Becke\cite{becke1993new}, 
based on a linear interpolation of the adiabatic connection formula. 
The present 50\% hybrid of CPBE and HF exchange (which we denote ``constrained hybrid'' - CHYB) 
is the self-interaction corrected analogue of the 
PBE50 functional\cite{bernard2012general} that was employed in time-dependent DFT 
and many-body perturbation theory 
methods. 
In DFT applications, the PBE0 hybrid functional\cite{perdew1996rationale,adamo1999toward} 
with 25\% HF exchange is a more widely used parameterisation.

\begin{figure}
\begin{centering}
\includegraphics[width=0.9\linewidth]{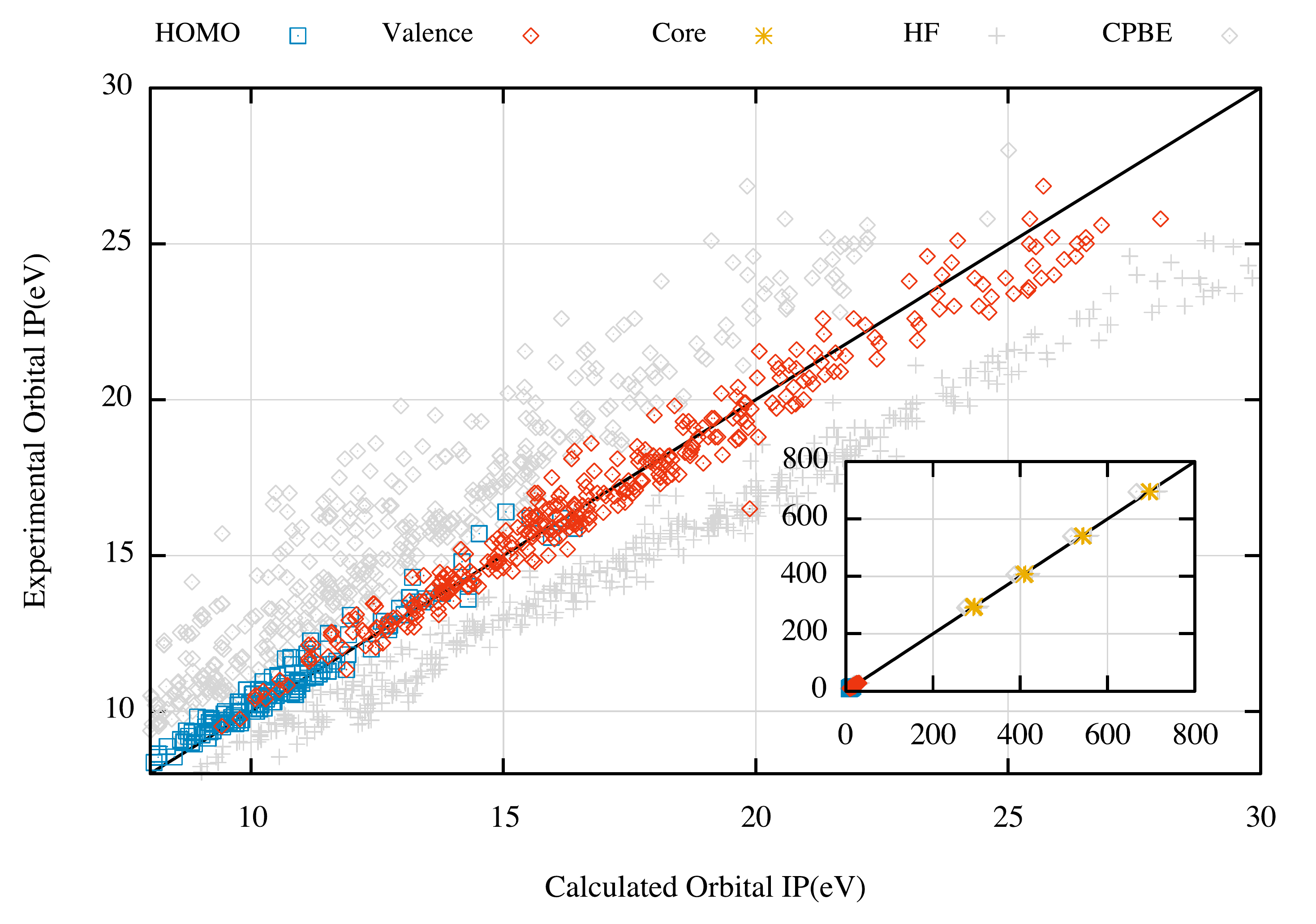}
\caption{Experimental IPs vs the IPs calculated orbital energies for the test set of molecules in 
Table~\ref{calc_expt_comp_table} using the cc-pVDZ basis set. Coloured points correspond to the hybrid method, grey points show results for CPBE (crosses) 
and HF(diamonds), the black line shows the ideal exact agreement between experimental and calculated results. Inset, shows the same results including high 
energy core orbitals.  } 
\end{centering}
\label{calc_expt_comp}
\end{figure}

The eigenvalue accuracy of the constrained hybrid method with $\alpha=0.5$ will now be investigated. 
Figure \ref{calc_expt_comp} demonstrates the agreement of the CHYB method with experimental results for the HOMO, valence and core orbitals. 
We see a clear improvement of the CHYB method over the CPBE and HF methods. 
Table \ref{calc_expt_comp_table} compares our new CHYB method with results from 
unconstrained and constrained PBE, HF and coupled-cluster singles and doubles(CCSD) 
results. 
These results demonstrate that the constrained 
hybrid method has a significantly reduced error over the other DFT methods for calculation of all IPs, and has accuracy 
approaching CCSD results. 
Ionisation energies of core orbitals show a significant improvement over PBE and HF results. These ionisation energies are largely independent 
of the surrounding chemical 
environment and as such the four groupings of points in Figure \ref{calc_expt_comp}(inset) correspond (from left to right) to C,N,O,F.
As can be seen from table \ref{calc_expt_comp_table}, the error for valence orbitals is in general positive; 
for core orbitals this error becomes negative. 
This is likely caused by the HF error increasing more relative to CPBE, due to the neglect of relaxation effects. 
The core orbitals especially are expected to have a large relaxation error and the HF error to increase.
Of the molecules investigated with this hybrid, outlying results are for the molecules O$_3$ with one valence orbital underestimated by 3eV and SiF$_4$ where every orbital 
has errors over 1eV. The large error on these molecules requires further investigation, although similarly poor results were also found for the CCSD calculations for 
O$_3$\cite{ranasinghe2019vertical}.
As would be expected from the imposition of additional constraints, the calculated total energy is higher than the unconstrained 50\% PBE hybrid and in 
line with the increase in energy when applying the constrained method\cite{pitts2018performance}. 
An increase in the total energy of $\sim0.1$meV was found for the Hydrogen Fluoride molecule. 
This small increase was maintained for a range of interatomic separations showing that the addition of the constrained method 
will have minimal impact on the energetic behaviour compared to the unconstrained method.   
This hybrid will therefore have the same energetic behaviour as the 50\% PBE hybrid method. 
\setlength{\tabcolsep}{6pt}
\begin{table*}
\begin{center}
\renewcommand{\arraystretch}{0.5}
\begin{tabular}{cccccccc}
\hline
\hline
&\multicolumn{3}{c}{All Valence States} &&\multicolumn{3}{c}{HOMO} \\
\cline{2-4} \cline{6-8}
&\#Orbs&AE&AAE&&\#Orbs&AE&AAE \\
\hline
cc-pVDZ\\
\hline
HF&381&2.90&2.90&&136&1.63&1.63\\
PBE&381&-5.02&5.02&&136&-4.41&4.41\\
CPBE&381&-2.84&2.84&&136&-2.06&2.06\\
CHYB&381&-0.01&0.52&&136&-0.24&0.34\\
IP-EOM-CCSD&381&-0.07&0.34&&136&-0.12&0.24\\
\hline
cc-pVTZ\\
\hline
HF&91&2.58&2.58&&28&1.62&1.62\\
PBE&91&-4.71&4.9&&28&-4.79&4.79\\
CPBE&91&-1.6&1.63&&28&-1.46&1.46\\
CHYB&91&0.48&0.54&&28&0.11&0.25\\
IP-EOM-CCSD&91&0.15&0.27&&28&-0.04&0.15\\
\hline
&\multicolumn{3}{c}{Core States}&&&&\\
\cline{2-4}
cc-pVTZ&\#Orbs&AE&AAE&&&&\\
\hline 
HF&15&17.34&17.34&&&&\\
PBE&15&-25.93&25.93&&&&\\
CPBE&15&-22.06&22.06&&&&\\
CHYB&15&-2.53&2.53&&&&\\
IP-EOM-CCSD&15&0.98&0.98&&&&\\
\hline 
\hline
\end{tabular} 
\caption{The errors in  calculated IPs for HOMO and valence orbitals for several methods, showing the number of orbitals involved (\#Orbs), the average error (AE), and the absolute average error (AAE). Results for core orbitals of the molecules in Table \ref{Correl_results} are also shown for the cc-pVTZ basis set. }
\end{center}
\label{calc_expt_comp_table} 
\end{table*}

Table \ref{All_orbs_final_results} shows the full orbital energies for a selection of molecules, compared with experimental results\cite{chong2002interpretation}. 
The agreement with experiment of the self-interaction corrected hybrid functional can clearly be seen throughout these molecules. Core orbitals have the largest 
absolute error due to their large eigenvalues, however even for these orbitals the constrained hybrid method has average errors smaller than 1eV.
The present method demonstrates a significant improvement in calculating the spectra of these molecules when compared to HF and the CPBE methods. 
\begin{table}
\begin{center}
\setlength{\tabcolsep}{5pt}
\renewcommand{\arraystretch}{0.5}
\begin{tabular}{cccccc}
\hline
\hline
MOL&Expt&HF&PBE&CPBE&CHYB\\
\hline
N$_\text{2}$&15.58&0.73&-5.45&-1.81&0.20\\
&16.93&0.21&-5.63&-1.97&-1.12\\
&18.75&2.71&-5.12&-1.47&1.05\\
&37.30&2.08&-9.58&-5.95&-1.11\\
&409.98&17.00&-26.57&-23.33&0.88\\
&409.98&17.08&-26.55&-23.30&0.94\\
\\
CO&14.01&1.13&-5.00&-1.41&0.10\\
&16.91&0.22&-5.30&-1.72&-0.56\\
&19.72&2.10&-5.65&-2.07&0.44\\
&38.30&2.69&-9.34&-5.77&-0.69\\
&296.21&13.17&-23.92&-20.73&-0.41\\
&542.55&19.81&-29.33&-26.30&1.35\\
\\
HF&16.19&1.29&-7.17&-1.83&0.06\\
&19.90&0.63&-7.01&-1.71&-0.27\\
&39.60&3.65&-10.20&-4.90&0.23\\
&694.23&21.09&-34.59&-28.42&1.27\\
\\
H$_\text{2}$O&12.62&1.11&-5.92&-1.53&0.06\\
&14.74&0.97&-5.90&-1.52&0.00\\
&18.55&0.60&-5.86&-1.48&-0.21\\
&32.20&4.30&-7.38&-2.99&1.38\\
&539.9&19.5&-29.71&-26.23&1.14\\
\hline
\hline
&&HF&PBE&CPBE&CHYB\\
ALL&AE&6.29&-12.91&-8.88&0.22\\
&AAE&6.29&12.91&8.88&0.64\\
\\
CORE&AE&17.94&-28.44&-24.72&0.86\\
&AAE&17.94&28.44&24.72&1.00\\
\\
VALENCE&AE&1.63&-6.7&-2.54&-0.03\\
&AAE&1.63&6.7&2.54&0.50\\
\\
HOMO&AE&1.06&-5.88&-1.65&0.11\\
&AAE&1.06&5.88&1.65&0.11\\
\hline
\hline
\end{tabular}  
\caption{ Errors for the full set of orbitals in the molecules N$_\text{2}$, CO, HF, H$_\text{2}$O for the cc-pVTZ basis set, comparing to experimental results. With average errors(AE) and average absolute errors (AAE) for the subsets of orbitals. } 
\end{center}
\label{All_orbs_final_results} 
\end{table}

\section{Conclusions} 

From the results presented in this paper it is clear that the constrained hybrid method offers significant improvements in the approximation of orbital 
energies not just for the HOMO but for the full range of occupied orbitals.
The errors in orbital energies, below 1eV for valence orbitals and within 1-2 eV for core orbitals, are surprisingly small considering the semi-local DFA we employ. 
We attribute the improvement to the hybrid character of the exchange potential, whose equally weighted components 
are either SI-free (non-local Fock term) or corrected from the main errors of SIs (constrained DFA exchange potential).  
The total energy of this constrained hybrid remains relatively consistent with the PBE and HF-plus-PBE-correlation 
methods, and is increased over an unconstrained hybrid method by $\sim0.1$meV. 
The accuracy of these results, especially for core orbitals suggest that there is a cancellation of errors between the exchange potential 
in the constrained minimisation method and HF. 
For orbitals with large ionisation energies there should be significant orbital relaxation effects, which nevertheless are accounted for in the orbital energy 
spectrum of the constrained hybrid method.
%
%
%
As we found in Ref.~\cite{pitts2018performance}, the IPs calculated from the constrained method is largely independent of the DFA. 
Therefore the constrained hybrid method will show similar improvements in the calculation of IPs for a range of DFAs using a mixing parameter of 50\%. 
This mixing is in line with a simple linear interpolation of the adiabatic connection and while further 
improvements for specific orbitals may be obtained for a highly 
optimised value of $\alpha$, the choice of $\alpha=0.5$ is expected to work for a variety of systems and orbitals. 
Finally, the accuracy of these results is consistent with 
recent arguments\cite{chong2002interpretation,hamel2002kohn,ranasinghe2019vertical,bartlett2019adventures} 
that the DFT Koopmans' theorem holds approximately for all orbitals, although to achieve the systematic high accuracy of 
the present results a hybrid exchange potential is necessary. 

\section*{Acknowledgments}
TCP and NIG acknowledge financial support from The Leverhulme Trust, through a Research Project Grant with number RPG-2016-005.  NIG thanks
Prof. Rod Bartlett for helpful discussions during his visit to Durham University in 2019 and acknowledges the Institute of Advanced Study at Durham
University for hosting this visit.\\

The data that support the findings of this study are available from the corresponding author upon reasonable request.



\end{document}